\documentclass[aps,prb,reprint]{revtex4-2}

\usepackage{epsf,graphicx,epstopdf}

\renewcommand{\vec}[1]{{\bf #1}}
\renewcommand{\d}{{\rm d}}
\renewcommand{\i}{{\rm i}}
\newcommand{\pard}{{\rm\partial}}
\newcommand{\e}{\mathop{\rm e}\nolimits}

\begin{document}

\title{Mie scattering theory: A review of physical features and limitations}

\author{Yuriy~A.~Akimov}
\email{akimov@ihpc.a-star.edu.sg}
\affiliation{Institute of High Performance Computing (IHPC), Agency for Science, Technology and Research (A*STAR),\\
1 Fusionopolis Way, \#16-16 Connexis, Singapore 138632}

\date{\today}

\begin{abstract} 
Mie theory is the classical problem for modeling of light scattering by spherical particles.    
In this paper, we perform a spherical harmonic analysis of its solution for the induced fields to reveal the physics underlying the resonant behavior predicted for scattering and absorption. 
We disclose the two distinct groups of current-sourced and current-free scattered fields, whose interference makes light-matter interaction resonant in Mie theory for every orbital index. 
Being a model, the current-free scattered fields naturally limit applicability of the classical solution. 
We discuss those limitations and demonstrate the ways for further refinement of the theory for description of the excitation source, sphere interface and scatterer localization. 
\end{abstract}

\maketitle

\section{Introduction}

Mie theory is the classical problem for light scattering on a spherical particle \cite{Stratton:2007,Jackson:1999,Bohren:1998}. 
It is named after German physicist Gustav Mie, who first published his paper on scattering of an electromagnetic wave on a sphere in 1908 \cite{Mie:1908}. 
Since then, this theory played a key part in shaping of our knowledge about light-matter interaction.
It bridged the theories of Rayleigh scattering for small particles and Rayleigh-Gans-Debye scattering for larger ones \cite{Strutt:1871}, enabling calculations for particles of arbitrary size. 
It explained frequency-selective scattering and size-dependent color change, which occur in solutions of colloid nanoparticles.
Later, the theory was successfully extended to multilayer spheres \cite{Aden:1951} and non-plane incident fields emitted by dipoles \cite{Ruppin:1982} or swift electrons \cite{deAbajo:1999}. 
Now, Mie theory is extensively used for design of optical resonances in plasmonic and photonic structures at the nanoscale \cite{Li:2014,Png:2017,Liu:2020,Menguc:2023}.

Despite the long history of success, Mie theory keeps a number of fundamental questions unanswered. 
For instance, why Mie scattering and absorption behave differently with change of particle size? 
Why does resonant scattering monotonically increase with particle size, while resonant absorption does not \cite{Akimov:2009}? 
Mie theory explains these behaviors by resonant nature of the Mie scattering coefficients. 
But how do they become resonant? 
What is the physics underlying their resonances? 
If the resonances in Mie coefficients arise from interference of the dipole and higher-order currents induced in the sphere as was suggested in Refs.~\cite{Grahn:2012,Miroshnichenko:2015,Luk'yanchuk:2017}, then how do similar resonances appear in deeply subwavelength metal particles \cite{Akimov:2012,Kolwas:2013} whose higher-order currents are negligible? 
How can a dipole current along make the Mie scattering coefficients resonant? 
Other questions arise from the way the light-matter interaction is treated in Mie theory. 
The theory successfully predicts static flows for the electromagnetic energy, but not for the momentum. 
This suggests that Mie theory provides a truncated description for light-matter interaction. 
Then, how limited is that description? 
All these questions highlight lack of our understanding of the Mie scattering.
To address them, we perform a rigorous spherical harmonic analysis of Mie theory revealing the physics underlying its solution for scattering and absorption of light.

\section{Mie theory}

Mie theory describes scattering and absorption of harmonic plane waves of angular frequency $\omega$ incident on a spherical particle. 
Following Maxwell's equations, electromagnetic fields in domains with uniform dielectric permittivity $\varepsilon$ in the absence of any external charges and currents can be completely described with the transverse magnetic (TM) and transverse electric (TE) fields as follows \cite{Li:2014,Png:2017}: 
\begin{eqnarray}
	&\displaystyle 
	\vec H=\left(\vec H^{\rm TM}-\frac{\i}{k_0} \sqrt{\frac{\varepsilon_0}{\mu_0}}\nabla\times \vec E^{\rm TE}\right)\e^{-\i\omega t},\label{H}\\
	&\displaystyle 
	\vec E=\left(\vec E^{\rm TE}+\frac{\i}{k_0\varepsilon}\sqrt{\frac{\mu_0}{\varepsilon_0}}\nabla\times \vec H^{\rm TM}\right)\e^{-\i\omega t},\label{E}
\end{eqnarray}
where $\vec H^{\rm TM}$ and $\vec E^{\rm TE}$ are the amplitudes of the governing fields for the TM and TE polarizations, $k_0=\omega\sqrt{\varepsilon_0\mu_0}$ is the vacuum wavenumber, $\varepsilon_0$ and $\mu_0$ are the electric and magnetic constants.
In the spherical geometry, the governing TM and TE fields can be decomposed over the vector spherical harmonics of different orbital and azimuthal indices $l, m$ to separate their radial and angular dependencies \cite{Li:2014,Png:2017}:
\begin{eqnarray}
	&\displaystyle 
	\vec H^{\rm TM}=\sum_{l=0}^\infty\sum_{m=-l}^l
{H}_{lm}\left(\vec e_\phi\frac{\pard Y_{lm}}{\pard \theta}-\frac{\vec e_\theta}{\sin \theta}\frac{\pard Y_{lm}}{\pard \phi}\right),\label{H_TM_lm}\\
	&\displaystyle 
	\vec E^{\rm TE}=\sum_{l=0}^\infty\sum_{m=-l}^l
{E}_{lm}\left(\vec e_\phi\frac{\pard Y_{lm}}{\pard \theta}-\frac{\vec e_\theta}{\sin \theta}\frac{\pard Y_{lm}}{\pard \phi}\right),\label{E_TE_lm}
\end{eqnarray}
where $(r,\theta,\phi)$ are the spherical coordinates, and $\vec e_r, \vec e_\theta, \vec e_\phi$ are the respective unit vectors.
In this separation, the angular dependence is fully given by the scalar spherical harmonics 
\begin{equation}
	Y_{lm}(\theta,\phi)=\sqrt{\frac{2l+1}{4\pi}\frac{(l-m)!}{(l+m)!}}\,
	P_l^m (\cos\theta)\e^{\i m\phi},
\end{equation}
where $P_l^m (\cos\theta)$ are the associated Legendre polynomials, while the radial dependence is given by the scalar functions ${H}_{lm}$ and ${E}_{lm}$ generally independent of each other and obeying the wave equations
\begin{eqnarray}
	&\displaystyle 
	\frac{\d^2 H_{lm}}{\d r^2}+\frac{2}{r}\frac{\d H_{lm}}{\d r}-\left[\frac{l(l+1)}{r^2}-k_0^2\varepsilon\right]H_{lm}=0, \label{WEd_H}\\
	&\displaystyle
		\frac{\d^2 E_{lm}}{\d r^2}+\frac{2}{r}\frac{\d E_{lm}}{\d r}-\left[\frac{l(l+1)}{r^2}-k_0^2\varepsilon\right]E_{lm}=0. \label{WEd_E}
\end{eqnarray}

For a spherical particle of radius $R$, Mie theory defines the radial functions in a piecewise manner \cite{Li:2014}:  
\begin{eqnarray}
	&\displaystyle 
	{E}_{lm}=\left\{
	\begin{array}{ll}
	E_{lm}^{\rm int}, & r<R,\\[3pt]
	E_{lm}^{\rm inc}+E_{lm}^{\rm sca}, & r>R,
	\end{array}
	\right.\\
	&\displaystyle 
	{H}_{lm}=\left\{
	\begin{array}{ll}
	H_{lm}^{\rm int}, & r<R,\\[3pt]
	H_{lm}^{\rm inc}+H_{lm}^{\rm sca}, & r>R,
	\end{array}
	\right.
\end{eqnarray}
where the incident, scattered and internal fields are given by
\begin{eqnarray}
	&\displaystyle 
	H_{lm}^{\rm inc}=\widetilde{H}_{lm} j_l(k_e r), \label{H_inc}\\
	&\displaystyle
	E_{lm}^{\rm inc}=\widetilde{E}_{lm} j_l(k_e r), \label{E_inc}\\
	&\displaystyle 
	H_{lm}^{\rm sca}=-a_{l} \widetilde{H}_{lm} h_l^{(1)}(k_e r), \label{H_sca}\\
	&\displaystyle 
	E_{lm}^{\rm sca}=-b_{l} \widetilde{E}_{lm} h_l^{(1)}(k_e r),\label{E_sca}\\
	&\displaystyle 
	H_{lm}^{\rm int}= d_{l} \widetilde{H}_{lm} j_l(k_i r),\label{H_int}\\
	&\displaystyle 
	E_{lm}^{\rm int}= c_{l} \widetilde{E}_{lm} j_l(k_i r)\label{E_int}
\end{eqnarray}
with $k_{i,e}=k_0\varepsilon_{i,e}^{1/2}$ being the wavenumbers of transverse fields in the internal ($r<R$) and external ($r>R$) homogeneous domains denoted with the indices $i$ and $e$, accordingly. The radial distributions of the scattered and internal fields in Mie theory are chosen to be the spherical Hankel functions of the first kind $h_l^{(1)}(k_e r)$ and the spherical Bessel functions $j_l(k_i r)$ providing finite solutions to Eqs.~(\ref{WEd_H}) and (\ref{WEd_E}) in the two domains, while the distributions of the incident fields are taken from the spherical decomposition of the $x$-polarized transverse electromagnetic (TEM) plane wave propagating in the $z$-direction: 
\begin{eqnarray}
	&\displaystyle 
	{\vec E}^{\rm inc}=
		{\vec e}_x
	E_0\e ^{\i (k_e r \cos\theta-\omega t)},
\\
	&\displaystyle 
	{\vec H}^{\rm inc}=\sqrt\frac{\varepsilon_0\varepsilon_e}{\mu_0}
		{\vec e}_y
	E_0\e^{\i (k_e r \cos\theta-\omega t)},
\end{eqnarray}
where $E_0$ is the complex amplitude of the incident electric field. If the external medium is fully transparent with ${\rm Im}~\varepsilon_e=0$, then the decomposition results in the following spherical harmonics amplitudes for the incident fields \cite{Li:2014}: 
\begin{eqnarray}
	&\displaystyle 
	\widetilde{H}_{lm}=-\i ^{l}\sqrt{\frac{\pi(2l+1)}{l(l+1)}}\sqrt\frac{\varepsilon_0\varepsilon_e}{\mu_0}m\delta_{m,\pm1}E_0,\\
	&\displaystyle
	\widetilde{E}_{lm}=-\i ^{l+1}\sqrt{\frac{\pi(2l+1)}{l(l+1)}}\delta_{m,\pm1}E_0.
\end{eqnarray}
These amplitudes are used in induced fields (\ref{H_sca})--(\ref{E_int}) for normalization together with dimensionless  coefficients $a_{l}$, $b_{l}$, $c_{l}$ and $d_{l}$, known as Mie coefficients. The latter are given by the boundary conditions implying the continuity of the tangential components of $\vec E^{\rm TE}$ and $\vec H^{\rm TM}$ on the sphere interface $r=R$ \cite{Stratton:2007,Jackson:1999,Bohren:1998}:
\begin{eqnarray}
	&\displaystyle 
	a_{l}=\frac{q_i\psi_l(q_i)\psi_l'(q_e)-q_e\psi_l(q_e)\psi_l'(q_i)}
	{q_i\psi_l(q_i)\xi_l'(q_e)-q_e\xi_l(q_e)\psi_l'(q_i)},
\label{S:a_l}\\
	&\displaystyle 
	b_{l}=\frac{q_e\psi_l(q_i)\psi_l'(q_e)-q_i\psi_l(q_e)\psi_l'(q_i)}
	{q_e\psi_l(q_i)\xi_l'(q_e)-q_i\xi_l(q_e)\psi_l'(q_i)},
\label{S:b_l}\\
	&\displaystyle 
	c_{l}=\frac{q_i\psi_l(q_e)\xi_l'(q_e)-q_i\xi_l(q_e)\psi_l'(q_e)}
	{q_e\psi_l(q_i)\xi_l'(q_e)-q_i\xi_l(q_e)\psi_l'(q_i)},
\label{S:c_l}\\
	&\displaystyle 
	d_{l}=\frac{q_i\psi_l(q_e)\xi_l'(q_e)-q_i\xi_l(q_e)\psi_l'(q_e)}
	{q_i\psi_l(q_i)\xi_l'(q_e)-q_e\xi_l(q_e)\psi_l'(q_i)},
\label{S:d_l}
\end{eqnarray}
where $\psi_l(q_i)=q_ij_l(q_i)$ and $\xi_l(q_e)=q_eh_l^{(1)}(q_e)$ are the Riccati--Bessel functions, with $q_i=k_i R$ and $q_e=k_e R$.

With the scattered fields determined in the spherical system, Mie theory enables integration of the power scattered and absorbed by the particle and calculates the total cross-sections of scattering and absorption for the incident plane waves \cite{Stratton:2007,Jackson:1999,Bohren:1998}: 
\begin{eqnarray}
	&\displaystyle 
	\sigma_{\rm sca}=
	\frac{2\pi}{k_e^2}\sum\limits_{l=1}^\infty 
	(2l+1)(|a_l|^2+|b_l|^2),\label{sigma_sca}
\\
	&\displaystyle 
	\sigma_{\rm abs}=
	\frac{2\pi}{k_e^2}\sum\limits_{l=1}^\infty 
	(2l+1)[{\rm Re}(a_l+b_l)-(|a_l|^2+|b_l|^2)],~~~\label{sigma_abs}
\end{eqnarray} 
where the contributions by the TM and TE spherical vector harmonics are represented with the scattering coefficients $a_l$ and $b_l$, accordingly. 

\section{Features}

\subsection{Two types of scattered fields}

\begin{figure}[t]
	\includegraphics[width=0.47\textwidth,clip,trim={0.3cm 17cm 0.2cm 0cm}]{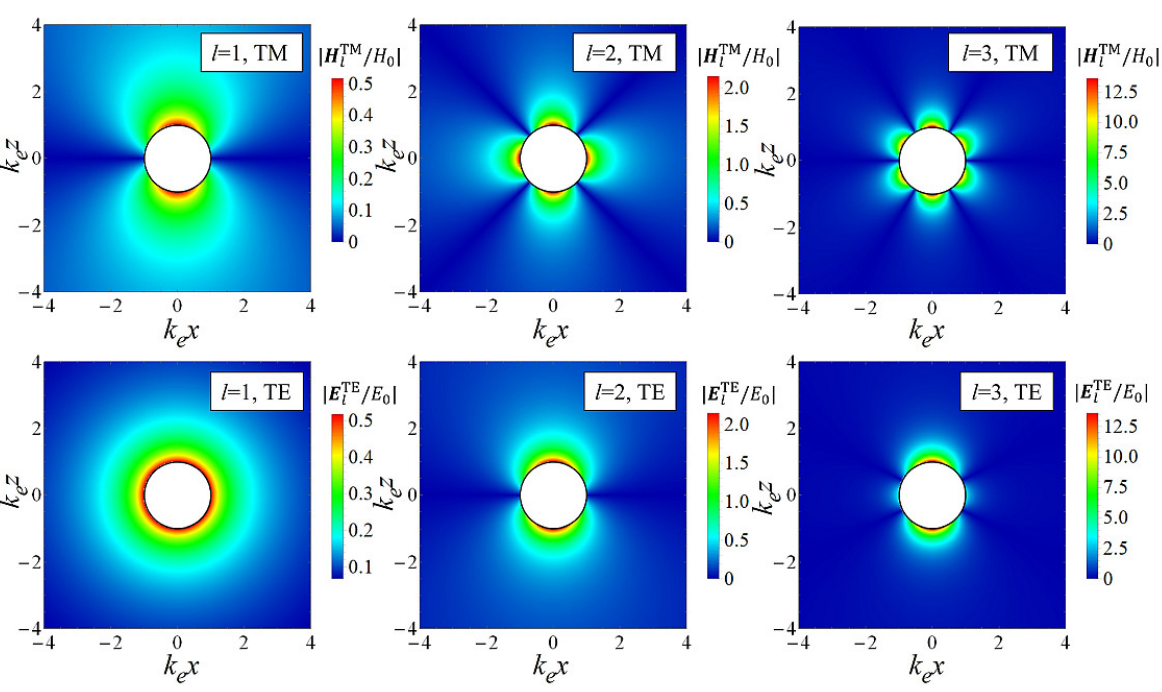}
	\caption{Contributions of source-free scattered fields of different orbital indices and polarizations to the governing fields $\vec H^{\rm TM}$ and $\vec E^{\rm TE}$ under $q_e=1$. The contributions are normalized by the amplitudes of the plane incident fields $E_0$ and $H_0=\sqrt{\varepsilon_0\varepsilon_e/\mu_0}E_0$.     
	\label{Fig1}}
\end{figure}

As Mie theory is linear to electromagnetic fields, one can split the incident fields into two parts: 
\begin{eqnarray}
	&\displaystyle 
	H_{lm}^{\rm inc}=H_{lm}^{\rm inc, 1}+H_{lm}^{\rm inc, 2}, \\
	&\displaystyle
	E_{lm}^{\rm inc}=E_{lm}^{\rm inc, 1}+E_{lm}^{\rm inc, 2},
\end{eqnarray}
and study how each part separately induces the scattered and internal fields:
\begin{eqnarray}
	&\displaystyle 
	H_{lm}^{\rm sca}=H_{lm}^{\rm sca, 1}+H_{lm}^{\rm sca, 2}, \\
	&\displaystyle
	E_{lm}^{\rm sca}=E_{lm}^{\rm sca, 1}+E_{lm}^{\rm sca, 2}, \\
	&\displaystyle 
	H_{lm}^{\rm int}=H_{lm}^{\rm int, 1}+H_{lm}^{\rm int, 2}, \\
	&\displaystyle
	E_{lm}^{\rm int}=E_{lm}^{\rm int, 1}+E_{lm}^{\rm int, 2}.
\end{eqnarray}
Particularly, one can decompose the incident fields into spherically outgoing and spherically incoming fields:
\begin{eqnarray}
	&\displaystyle 
	H_{lm}^{{\rm inc},j}=\frac{1}{2}\widetilde{H}_{lm}h_l^{(j)}(k_e r), \\
	&\displaystyle
	E_{lm}^{{\rm inc},j}=\frac{1}{2}\widetilde{E}_{lm}h_l^{(j)}(k_e r),
\end{eqnarray}
where $h_l^{(j)}(k_e r)$ are the spherical Hankel functions of the kind $j=1,2$.
Then, the respective internal and scattered fields can be written as follows:
\begin{eqnarray}
	&\displaystyle 
	H_{lm}^{{\rm sca},j}=-a_{lm}^{(j)}\widetilde{H}_{lm} h_l^{(1)}(k_e r), \\
	&\displaystyle 
	E_{lm}^{{\rm sca},j}=-b_{lm}^{(j)} \widetilde{E}_{lm} h_l^{(1)}(k_e r),\\
	&\displaystyle 
	H_{lm}^{{\rm int},j}= d_{lm}^{(j)} \widetilde{H}_{lm} j_l(k_i r),\\
	&\displaystyle 
	E_{lm}^{{\rm int},j}= c_{lm}^{(j)} \widetilde{E}_{lm} j_l(k_i r).
\end{eqnarray}
The new Mie coefficients appear to be
\begin{eqnarray}
	&\displaystyle 
	a_{l}^{(1)}=
	b_{l}^{(1)}=\frac{1}{2},
	\quad c_{l}^{(1)}=
	d_{l}^{(1)}=0
\label{S:cd_l1}
\end{eqnarray}
for the spherically outgoing incident fields and 
\begin{eqnarray}
	&\displaystyle 
	a_{l}^{(2)}=\frac{1}{2}\frac{q_i\psi_l(q_i)\zeta_l'(q_e)-q_e\zeta_l(q_e)\psi_l'(q_i)}
	{q_i\psi_l(q_i)\xi_l'(q_e)-q_e\xi_l(q_e)\psi_l'(q_i)},
\label{S:a_l2}\\
	&\displaystyle 
	b_{l}^{(2)}=\frac{1}{2}\frac{q_e\psi_l(q_i)\zeta_l'(q_e)-q_i\zeta_l(q_e)\psi_l'(q_i)}
	{q_e\psi_l(q_i)\xi_l'(q_e)-q_i\xi_l(q_e)\psi_l'(q_i)},
\label{S:b_l2}\\
	&\displaystyle 
	c_{l}^{(2)}=\frac{1}{2}\frac{q_i\psi_l(q_e)\zeta_l'(q_e)-q_i\zeta_l(q_e)\psi_l'(q_e)}
	{q_e\psi_l(q_i)\xi_l'(q_e)-q_i\xi_l(q_e)\psi_l'(q_i)},
\label{S:c_l2}\\
	&\displaystyle 
	d_{l}^{(2)}=\frac{1}{2}\frac{q_i\zeta_l(q_e)\xi_l'(q_e)-q_i\xi_l(q_e)\zeta_l'(q_e)}
	{q_i\psi_l(q_i)\xi_l'(q_e)-q_e\xi_l(q_e)\psi_l'(q_i)}
\label{S:d_l2}
\end{eqnarray}
for the spherically incoming incident fields, where $\zeta_l(q_e)=q_eh_l^{(2)}(q_e)$ are the spherically incoming Riccati-Bessel functions.

One can see the unusual scattering of spherically outgoing fields. Their scattered fields appear fully compensating the incident fields, $H_{lm}^{\rm sca, 1}=-H_{lm}^{\rm inc, 1}$, $E_{lm}^{\rm sca, 1}=-E_{lm}^{\rm inc, 1}$, and accompanied by the zero internal fields, $H_{lm}^{\rm int, 1}=0$, $E_{lm}^{\rm int, 1}=0$. This type of scattering corresponds to the {\it trivial} solution of Maxwell's equations with the zero total fields, $\vec E=0$, $\vec H=0$, inside and outside of the sphere. Such a solution exhibit nonzero scattered fields under the zero internal current induced in the scatterer. Thus, the scattered fields $H_{lm}^{{\rm sca},1}$ and $E_{lm}^{{\rm sca},1}$ are {\it source-free}. In contrast to them, the scattered fields $H_{lm}^{{\rm sca},2}$ and $E_{lm}^{{\rm sca},2}$ are accompanied by nonzero internal fields, $H_{lm}^{\rm int, 1}\neq0$, $E_{lm}^{\rm int, 1}\neq0$, highlighting their {\it current-sourced} nature. In other words, Mie theory deals with two fundamentally different scattered fields: (i)  source-free and (ii) current-sourced ones.

\begin{figure}[t]
	\includegraphics[width=0.47\textwidth,clip,trim={0.5cm 1.5cm 0cm 2cm}]{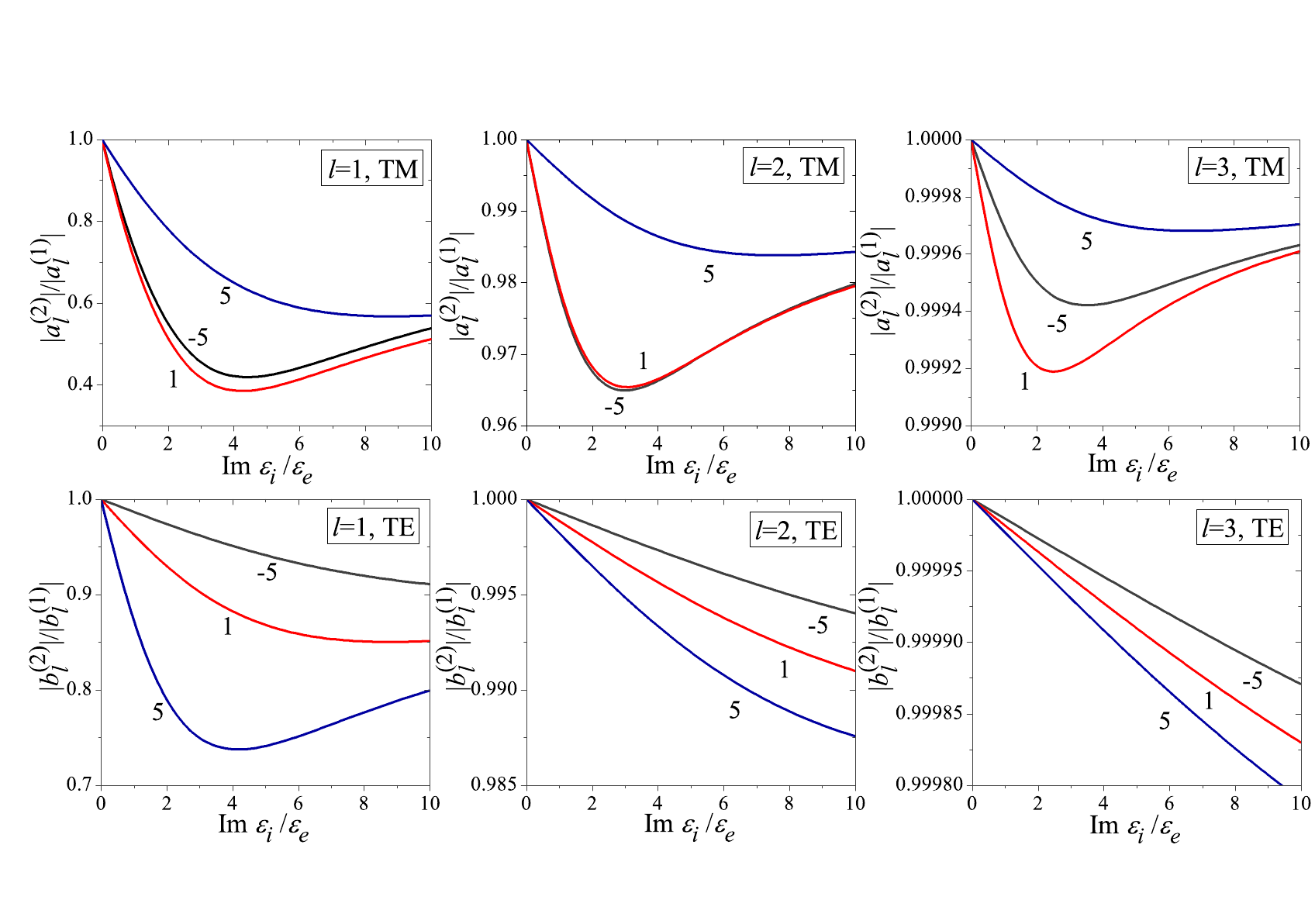}
	\caption{Amplitude ratios of the current-sourced and source-free scattered fields as functions of ${\rm Im}~\varepsilon_i/\varepsilon_e$ for different orbital indices and polarizations under $q_e=1$. The labels show the values of ${\rm Re}~\varepsilon_i/\varepsilon_e=-5$, 1 and 5 kept fixed for the respective curves.     
	\label{Fig2}}
\end{figure}

The paradox of two kinds of scattered fields comes from the formulation of Mie theory. 
It extends the conventional electrodynamic formulation of current-sourced fields and introduces free sourceless fields. 
Namely, the incident fields used in Mie theory are source-free \cite{Stratton:2007,Jackson:1999,Bohren:1998}. 
As a result, all the fields in Mie theory, including the scattered ones, are composed of the current-sourced and source-free electromagnetic fields. The former are excited by the internal currents induced in the scatterer, while the latter appear as the trivial response of Maxwell's equations to the introduced free incident fields. 

Noteworthy that the source-free scattered fields are independent of the sphere size and material, so the effect of scatterer in Mie theory is solely defined by the current-sourced fields. The typical distribution of the source-free scattered fields with different orbital indices $l$ is shown in Fig.~\ref{Fig1}. The current-sourced scattered fields have similar distributions, but with different complex amplitudes. For non-absorbing particles with ${\rm Im}~
\varepsilon_i=0$, the amplitudes of the current-sourced scattered fields differ from those of the source-free scattered fields by phase only: $|a_l^{(2)}|=|a_l^{(1)}|$ for the TM polarization and $|b_l^{(2)}|=|b_l^{(1)}|$ for the TE one. For absorbing particles with ${\rm Im}~\varepsilon_i>0$, the complex amplitudes of the source-free and current-sourced scattered fields differ by both amplitudes and phases, as shown in Fig.~\ref{Fig2}.

\subsection{Super-radiating and non-radiating states}

\begin{figure}[t]
	\includegraphics[width=0.48\textwidth,clip,trim={0.4cm 17cm 0.4cm 0cm}]{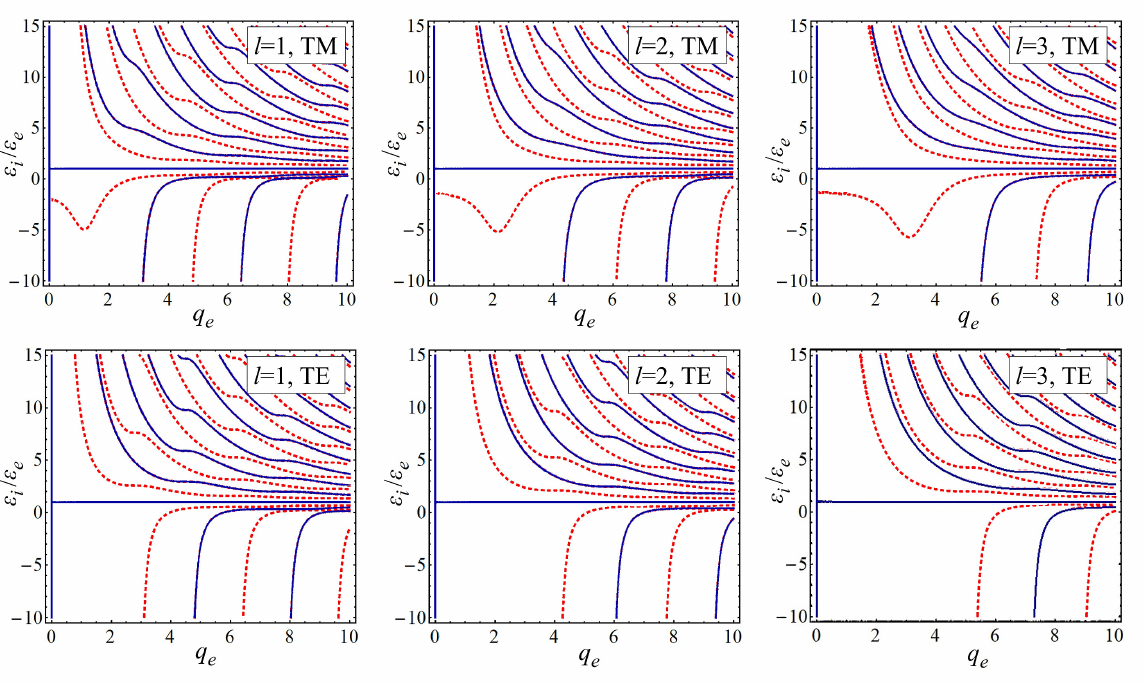}
	\caption{Super-radiating (dashed lines) and non-radiating (solid lines) states of the TM and TE polarizations for the orbital indices $l=1,2,3$. 
	\label{Fig3}}
\end{figure}

\begin{figure}[b]
	\includegraphics[width=0.48\textwidth,clip,trim={1.5cm 1.3cm 0.2cm 2.5cm}]{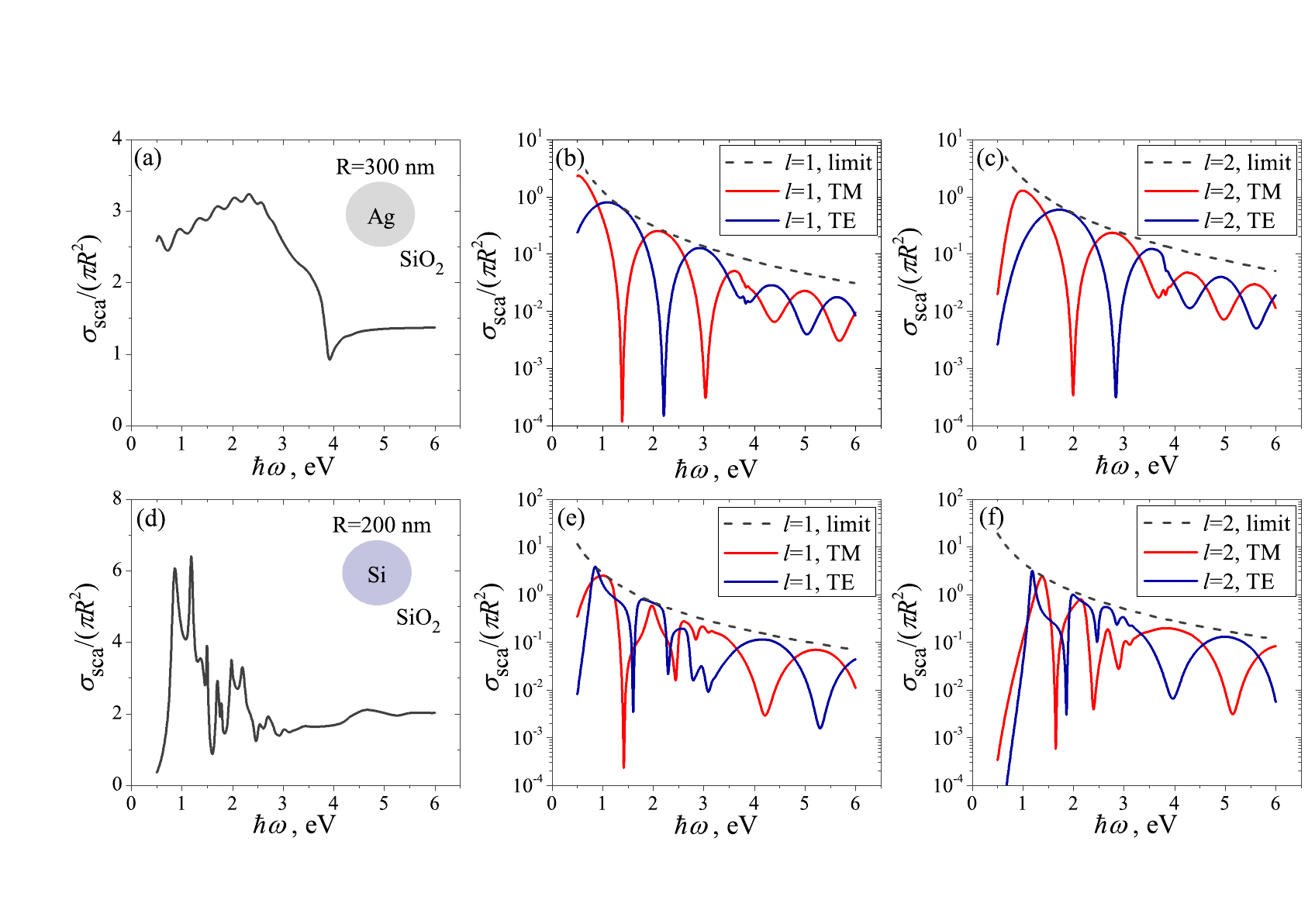}
	\caption{Normalized scattering cross-section with respective contributions of the TM and TE fields with $l=1,2$ for (a)--(c) a silver particle of $R=300$ nm and (d)--(f) a silicon particle of $R=200$ nm embedded in silicon dioxide. The limits in (b), (c) and (e), (f) are given by $\sigma_{{\rm sca},l}^{\rm sr}$.
	\label{Fig4}}
\end{figure}

As the scattered fields in Mie theory are not fully current-sourced, their effect on $\sigma_{\rm sca}$ and $\sigma_{\rm abs}$ can be understood from the interference of the two types of fields:
\begin{eqnarray}
	&\displaystyle 
	a_{l}=a_{l}^{(1)}+a_{l}^{(2)},\\
	&\displaystyle 
	b_{l}=b_{l}^{(1)}+b_{l}^{(2)},
\end{eqnarray}
rather than from the classical analysis of the current induced in the particle \cite{Jackson:1999,Grahn:2012} that is obviously inapplicable to the source-free electromagnetic fields. 
Following this interference, spherical particles support a number of {\it super-radiating} and {\it non-radiating states} for each polarization and orbital index $l$, as shown in Fig.~\ref{Fig3}. 

According to Eq.~(\ref{sigma_sca}), the super-radiating states exhibit either $a_{l}=1$ or $b_{l}=1$. They appear at those size parameters $q_e$ and relative dielectric permittivities $\varepsilon_i/\varepsilon_e$, which result in the constructive interference of the source-free and current-sourced scattered fields: $a_{l}^{(2)}=a_{l}^{(1)}$ for the TM polarization and $b_{l}^{(2)}=b_{l}^{(1)}$ for the TE one. Contrary, the non-radiating states, also known as anapoles, exhibit $a_{l}=0$ or $b_{l}=0$ caused by the destructive interference of the two kinds of scattered fields: $a_{l}^{(2)}=-a_{l}^{(1)}$ for the TM polarization and $b_{l}^{(2)}=-b_{l}^{(1)}$ for the TE one.

As the super-radiating and non-radiating states require purely real dielectric permittivity, those states are not achievable for real materials with finite frequency dispersion and dissipation. Nonetheless, they define the limits, when every polarization and orbital index contribute
\begin{eqnarray}
	&\displaystyle\sigma_{l}^{\rm sca,sr}=\frac{2\pi}{k_e^2}(2l+1), \quad \sigma_{{\rm abs},l}^{\rm sr}=0,\\
		&\sigma_{{\rm sca},l}^{\rm nr}=0, \qquad \sigma_{{\rm abs},l}^{\rm nr}=0
\end{eqnarray} 
to the total cross-sections $\sigma_{\rm sca}$ and $\sigma_{\rm abs}$. Figure~\ref{Fig4} shows the typical interference fringes in scattering cross-section of silver and silicon spheres with the upper limits given by $\sigma_{{\rm sca},l}^{\rm sr}$. 

\begin{figure}[b]
	\includegraphics[width=0.48\textwidth,clip,trim={1.5cm 1.3cm 0.2cm 2.5cm}]{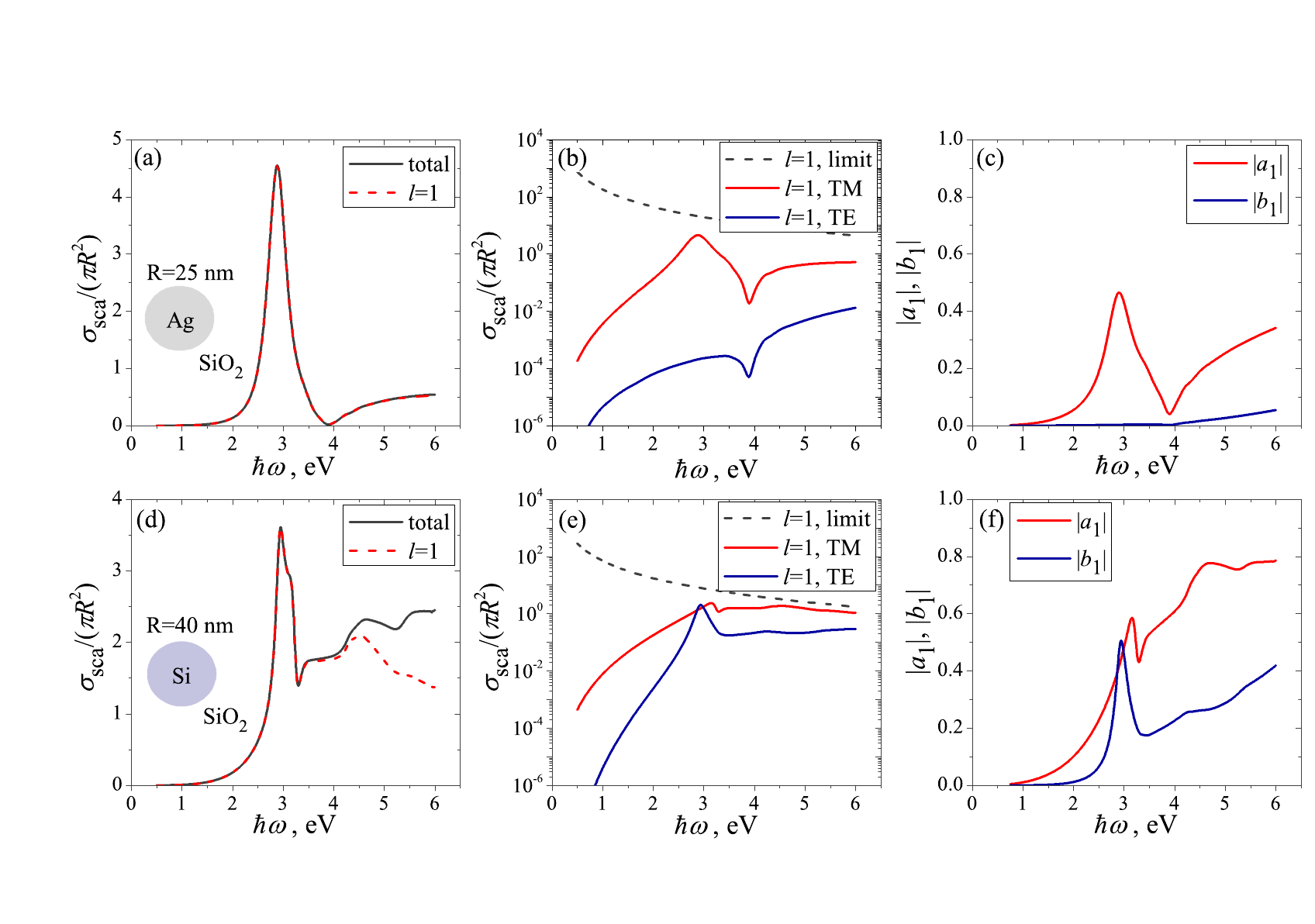}
	\caption{Normalized scattering cross-section with respective contributions of the TM and TE fields with $l=1,2$ for (a)--(c) a silver particle of $R=25$ nm and (d)--(f) a silicon particle of $R=40$ nm embedded in silicon dioxide. The limits in (b) and (e) are given by $\sigma_{{\rm sca},l}^{\rm sr}$.
	\label{Fig5}}
\end{figure}

Noteworthy that the density of interference fringes observed in scattering spectrum is heavily affected by the size of spheres. Figure~\ref{Fig3} suggests that by moving towards lower $q_e$, we can reduce the number of available supper-radiating states and, hence, the density of scattering fringes for every $l$. Furthermore, if we go down to $q_e\leq 2$, we can filter out contributions with high orbital indices, whose current-sourced and source-free scattered fields can be kept destructively interfering each other. Such cases are shown in Fig.~\ref{Fig5} for silver and silicon nanospheres with the dominant dipolar ($l=1$) contributions. They demonstrate one TM maximum for the metallic particle and two TE and one TM maxima for the dielectric one, which arise from the interference of the dipolar scattered fields that runs in a more constructive way. 

\begin{figure}[t]
	\includegraphics[width=0.48\textwidth,clip,trim={0.2cm 17cm 0.4cm 0cm}]{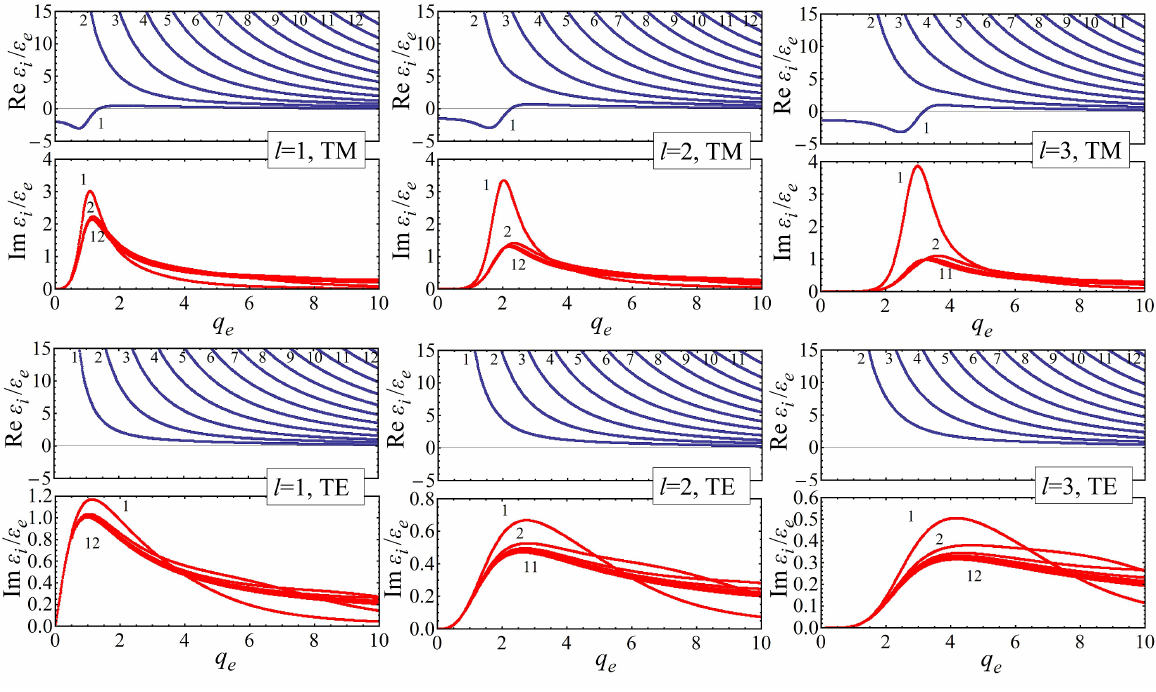}
	\caption{Super-absorbing states of the TM and TE polarizations for the orbital indices $l=1,2,3$. 
	\label{Fig6}}
\end{figure}

\begin{figure}[b]
	\includegraphics[width=0.48\textwidth,clip,trim={1.2cm 1.3cm 0.2cm 2.8cm}]{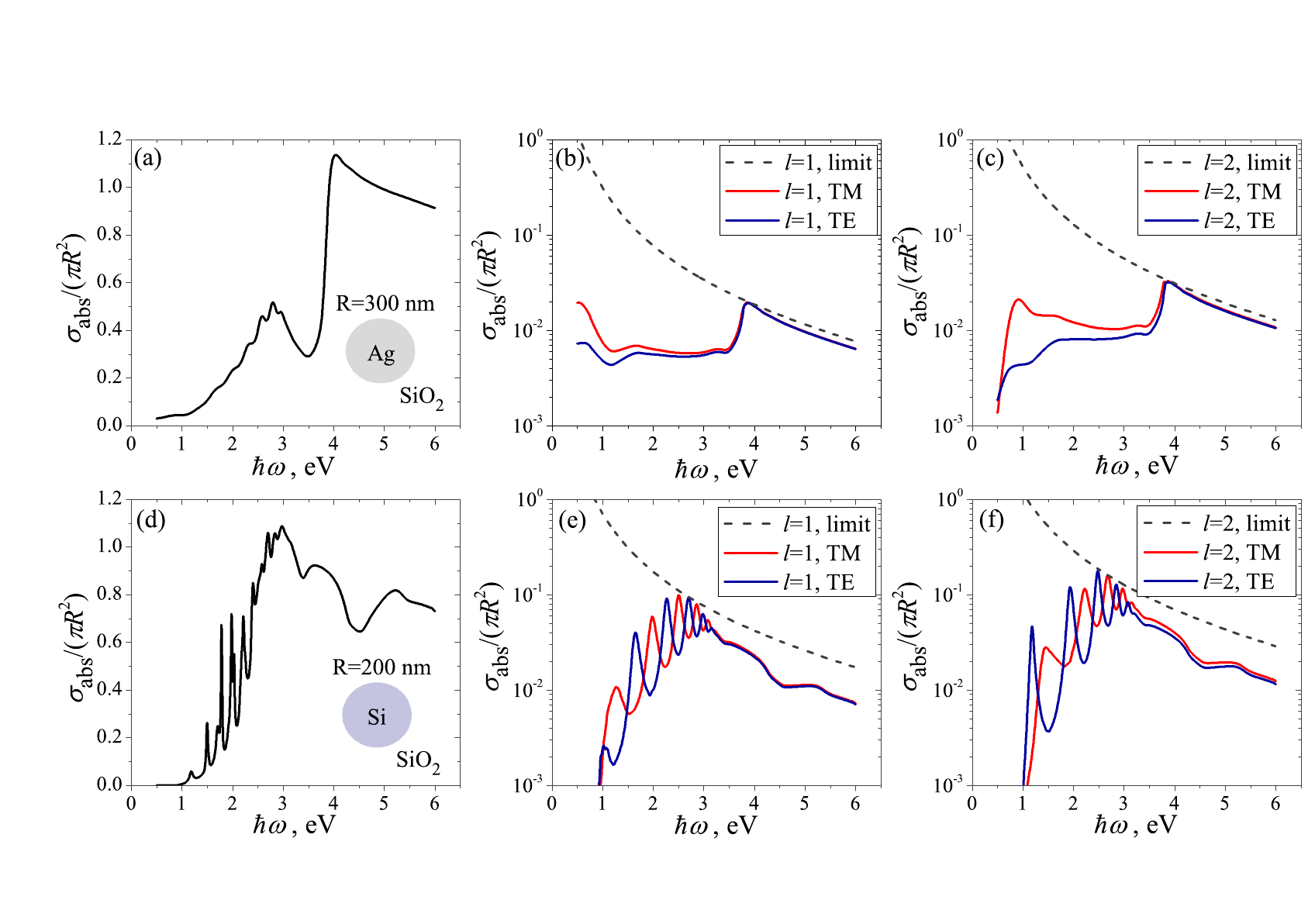}
	\caption{Normalized absorption cross-section with respective contributions of the TM and TE fields with $l=1,2$ for (a)--(c) a silver particle of $R=300$ nm and (d)--(f) a silicon particle of $R=200$ nm embedded in silicon dioxide. The limits in (b), (c) and (e), (f) are given by $\sigma_{{\rm abs},l}^{\rm sa}$.
	\label{Fig7}}
\end{figure}

In addition to the mentioned maxima, Fig.~\ref{Fig5} demonstrates several scattering minima caused by the interference of the dipolar scattered fields that runs in a more destructive way. The low-frequency minimum common to both the silver and silicon cases arises from the fundamental non-radiating state given by $q_e=0$. This state is independent of the polarization, orbital index and scatterer material, as shown in Fig.~\ref{Fig3}. Therefore, the associated regime, known as the {\it Rayleigh scattering} \cite{Strutt:1871,Bohren:1998}, is common to all structures. It is observed for $q_e\ll1$ with $|a_1|\gg|b_1|,|a_2|,|b_2|,...$, where $|a_1|\propto q_e^3$. The effect of another fundamental non-radiating state given by $\varepsilon_i=\varepsilon_e$ is seen in Fig.~\ref{Fig5}(a) for the silver particle around $\hbar\omega=3.9$ eV. This state is independent of the polarization, orbital index and scatterer size, as shown in Fig.~\ref{Fig3}, but requires a real-valued $\varepsilon_i$. As a result, the associated minimum of scattering is common to the structures that exhibit low optical loss \cite{Akimov:2012,Kolwas:2013}.         

\subsection{Super-absorbing states}

Another limiting case important for understanding of Mie theory is the maximum achievable absorption. Following Eq.~(\ref{sigma_abs}), the {\it super-absorbing states} must exhibit either $a_{l}=1/2$ or $b_{l}=1/2$. These states require the complex values of $\varepsilon_i/\varepsilon_e$ optimized for every size parameter $q_e$, as shown in Fig.~\ref{Fig6}. The above conditions also can be considered as those when the current-sourced scattered fields vanish: $a_{l}^{(2)}=0$ for the TM polarization or $b_{l}^{(2)}=0$ for the TE one. In other words, the super-absorbing states appear exactly the {\it source-free scattering states}. Contrary to the super-scattering states, there are multiple TM and TE super-absorbing states with ${\rm Re~}\varepsilon_i/\varepsilon_e>0$ and only one TM state with ${\rm Re~}\varepsilon_i/\varepsilon_e<0$.

The super-absorbing states define the upper limits for absorption contributions
\begin{eqnarray}
	\sigma_{{\rm abs},l}^{\rm sa}=\frac{\pi}{2k_e^2}(2l+1)
\end{eqnarray} 
by every polarization and orbital index. At those states, the contribution to the scattering and absorption cross-section appears exactly the same, equal to the quarter of the super-radiating limit:  
\begin{eqnarray}
	\sigma_{{\rm sca},l}^{\rm sa}=\sigma_{{\rm abs},l}^{\rm sa}=\frac{1}{4}\sigma_{{\rm sca},l}^{\rm sr}.
\end{eqnarray}

\begin{figure}[b]
	\includegraphics[width=0.48\textwidth,clip,trim={1.5cm 1.3cm 0.2cm 2.5cm}]{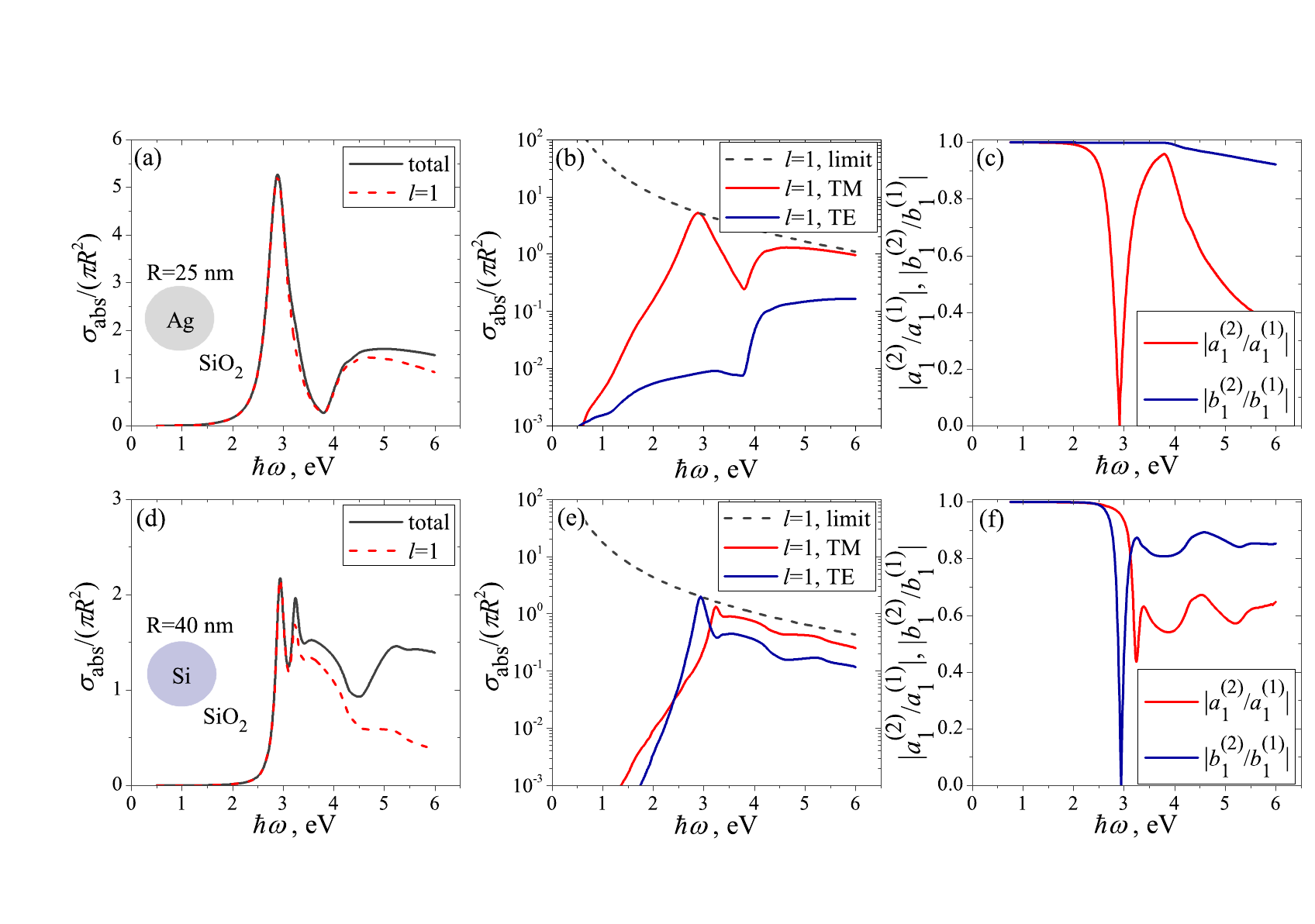}
	\caption{Normalized absorption cross-section with respective contributions of the TM and TE fields with $l=1,2$ for (a)--(c) a silver particle of $R=25$ nm and (d)--(f) a silicon particle of $R=40$ nm embedded in silicon dioxide. The limits in (b) and (e) are given by $\sigma_{{\rm abs},l}^{\rm sa}$.
	\label{Fig8}}
\end{figure}

Contrary to the super-radiating states, the super-absorbing ones are achievable for dispersive materials with finite dissipation under the proper size and material design. The typical absorption fringes given by the interference of the source-free and current-sourced scattered fields and bounded by $\sigma_{{\rm abs},l}^{\rm sa}$ are shown in Fig.~\ref{Fig7} for silver and silicon particles.   

Similar to scattering, the fringes in absorption are strongly affected by the size of spheres. We can reduce the density of fringes for every orbital $l$ by moving towards lower $q_e$, as suggested by Fig.~\ref{Fig6}. If we go down to $q_e\leq 2$, we can filter out contributions with high orbital indices, whose current-sourced and source-free scattered fields destructively interfere each other. Such cases are shown in Fig.~\ref{Fig8} for silver and silicon nanospheres with the dominant dipolar ($l=1$) contributions, where the respective TM and TE contributions are maximized for the particle absorption. Note that at the limiting cases given by $\sigma_{{\rm abs},l}^{\rm sa}$ the corresponding scattered fields are purely source-free.    

\section{Limitations}

In our analysis of the solution for induced fields, we already saw the modeling essence of Mie theory. Below, we will consider the major assumptions made in the theory for description of the excitation source, sphere interface and scatterer localization and discuss the ways to relax the limitations.

\subsection{Excitation source}

\begin{figure}[t]
	\includegraphics[width=0.47\textwidth,clip,trim={0.2cm 17cm 0.2cm 0.2cm}]{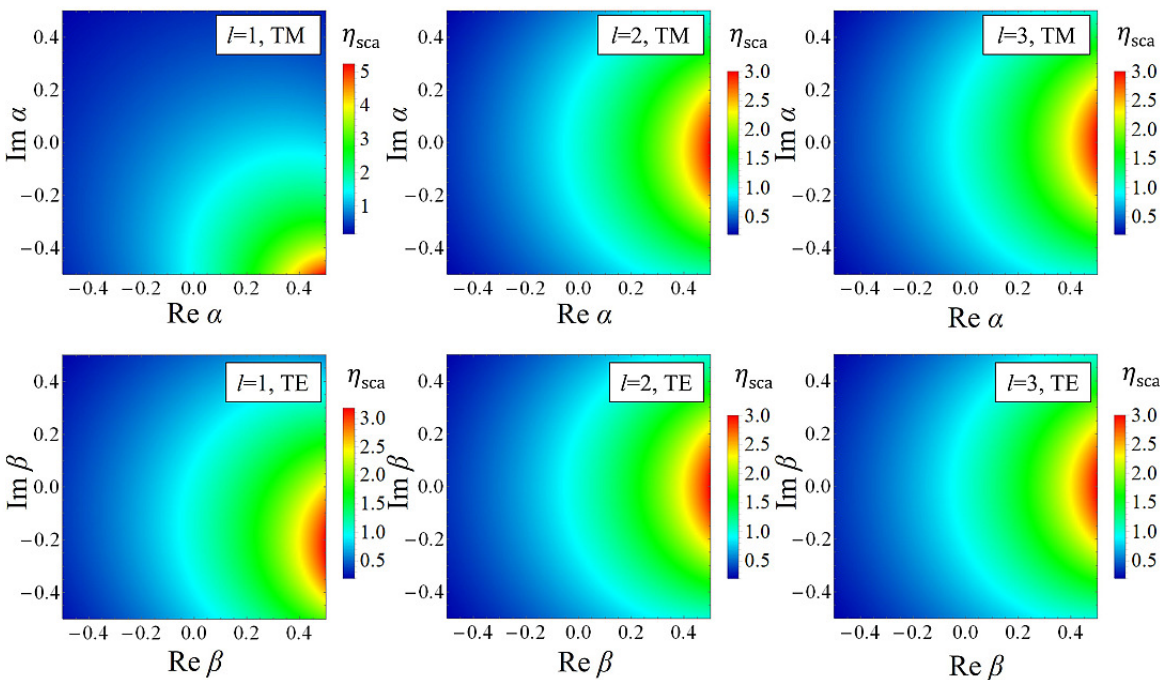}
	\caption{Ratios $\eta_{\rm sca}$ of the modified scattering cross-section contributions given by Eq.~(\ref{S_sca_source}) to the classical ones for the TM and TE scattered fields with $l=1,2,3$ under $q_e=1$ and $\varepsilon_i/\varepsilon_e=5+\i~0.1$.
	\label{Fig9}}
\end{figure}

Use of external incident fields for modeling of source of excitation is a common approach employed in many scattering problems \cite{Bohren:1998}, including Mie theory. It allows us to exclude the excitation current from consideration and truncate the problem to a smaller domain. At the same time, it makes the description of electromagnetic interaction between the excitation current and the scatterer electrodynamically incomplete. Indeed, incident fields (\ref{H_inc}) and (\ref{E_inc}) can be treated as the far-fields of an external current located at $r\geq R_{\rm ext}$ far from the scatterer, $R_{\rm ext}\gg R$, whereas the scattered fields written in the form of Eqs.~(\ref{H_sca}) and (\ref{E_sca}) do not support existence of any current at $r>R$. To account for the external excitation current, the scattered fields must be given by linear combinations of $h^{(1)}(k_e r)$ and $h^{(2)}(k_e r)$: 
\begin{eqnarray}
	&\displaystyle 
	H_{lm}^{\rm sca}=-a_{l}\widetilde{H}_{lm}[h_l^{(1)}(k_e r)+\alpha_l h_l^{(2)}(k_e r)], \\
	&\displaystyle 
	E_{lm}^{\rm sca}=-b_{l}\widetilde{E}_{lm}[h_l^{(1)}(k_e r)+\beta_l h_l^{(2)}(k_e r)]
\end{eqnarray}
for $R<r<R_{\rm ext}$, where $\alpha_l$ and $\beta_l$ are the complex interaction coefficients given by the source currents located at $r\geq R_{\rm ext}$. With this modification, Mie scattering coefficients change to
\begin{eqnarray}
	&\displaystyle 
	a_{l}=\frac{q_i\psi_l(q_i)\psi_l'(q_e)-q_e\psi_l(q_e)\psi_l'(q_i)}
	{q_i \psi_l(q_i) A_l'(q_e)-q_e A_l(q_e)\psi_l'(q_i)},
\label{a_l_ext}\\
	&\displaystyle 
	b_{l}=\frac{q_e\psi_l(q_i)\psi_l'(q_e)-q_i\psi_l(q_e)\psi_l'(q_i)}
	{q_e\psi_l(q_i) B_l'(q_e)-q_i B_l(q_e)\psi_l'(q_i)},
\label{b_l_ext}
\end{eqnarray}
where
\begin{figure}[t]
	\includegraphics[width=0.48\textwidth,clip,trim={0.2cm 17.3cm 0.2cm 0.2cm}]{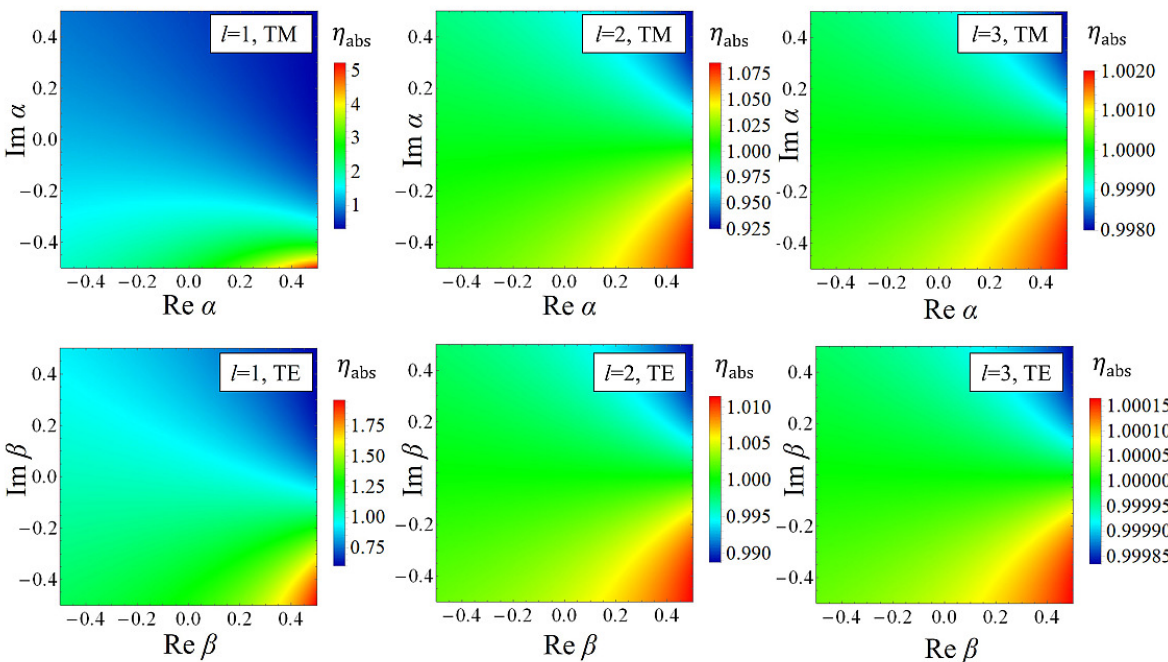}
	\caption{Ratios $\eta_{\rm abs}$ of the modified absorption cross-section contributions given by Eq.~(\ref{S_abs_source}) to the classical ones for the TM and TE scattered fields with $l=1,2,3$ under $q_e=1$ and $\varepsilon_i/\varepsilon_e=5+\i~0.1$.
	\label{Fig10}}
\end{figure}
\begin{eqnarray}
	&\displaystyle 
	A_l(q_e)=\xi_l(q_e)+\alpha_l \zeta_l(q_e),\\
	&\displaystyle 
	B_l(q_e)=\xi_l(q_e)+\beta_l \zeta_l(q_e).
\end{eqnarray}
Finally, the scattering and absorption cross-sections appear 
\begin{eqnarray}
	&\displaystyle 
	\sigma_{\rm sca}=
	\frac{2\pi}{k_e^2}\sum\limits_{l=1}^\infty 
	(2l+1)[|a_l|^2(1-|\alpha_l|^2)+|b_l|^2(1-|\beta_l|^2)],\nonumber
\\&\label{S_sca_source}\\
	&\displaystyle 
	\sigma_{\rm abs}=
	\frac{2\pi}{k_e^2}\sum\limits_{l=1}^\infty 
	(2l+1){\rm Re}[a_l(1-\alpha_l)+b_l(1-\beta_l)-\nonumber\\
	&\displaystyle 
	|a_l|^2(1-|\alpha_l|^2)-|b_l|^2(1-|\beta_l|^2)].\label{S_abs_source}
\end{eqnarray} 
As we can see, the account for external current changes the super-radiating, non-radiating and super-absorbing states and eventually affects the total scattering and absorption cross-sections, especially for low orbital indices, as shown in Figs.~\ref{Fig9} and \ref{Fig10}. 
 
\subsection{Sphere interface}

Although the original Mie theory focuses of energy characteristics of the source-scatterer interaction, it can be used to study momentum characteristics as well.
For instance, we can use Mie theory for calculation of the optical force that acts on the sphere from the source side:
\begin{eqnarray}
	&\displaystyle 
	\vec F^{\rm opt}=\frac{1}{2}{\rm Re}\int (\rho^*\vec E^{\rm inc}+\mu_0\vec J^*\times\vec H^{\rm inc})\d V.\label{F_opt}
\end{eqnarray} 
Here, $\rho$ is the volume density of the surface charge induced on the sphere interface:
\begin{eqnarray}
	&\displaystyle 
	\rho=\varepsilon_0\nabla\cdot\vec E=\widetilde\rho(\theta,\phi)\delta(r-R)\exp(-\i\omega t)
\end{eqnarray} 
with $\widetilde\rho(\theta,\phi)$ being the angular distribution and $\delta(r-R)$ being the radial Dirac delta-function distribution. The density of induced current $\vec J$ in Eq.~(\ref{F_opt}) is given by  
\begin{eqnarray}
	&\displaystyle
	\vec J=-\i\omega\varepsilon_0(\varepsilon-1)\vec E.
\end{eqnarray} 
The optical force is widely used for modeling of particle dynamics in colloid solutions. However, when it comes to calculation of the total force acting on the sphere: 
\begin{eqnarray}
	&\displaystyle 
	\vec F^{\rm tot}=\frac{1}{2}{\rm Re}\int (\rho^*\vec E+\mu_0\vec J^*\times\vec H)\d V,
\end{eqnarray} 
or the compression force of the surface charges:
\begin{eqnarray}
	\vec F^{\rm com}=\vec F^{\rm tot}-\vec F^{\rm opt},
\end{eqnarray} 
Mie theory naturally fails, as its solution for electric field $\vec E$ is discontinuous at $r=R$ and, hence, undefined at the location of the induced surface charge. 
It makes calculation of the Coulomb force in $\vec F^{\rm tot}$ mathematically impossible.
To resolve this issue, a continuous transition layer with radially varying $\varepsilon(r)$ between the sphere and the embedding medium is required. 
Introduction of such a layer disperses the induced charge and makes it volume, while the electric field distribution becomes continuous. 
This enables description of the forces acting on the sphere and other characteristics of momentum transfer.  

Obviously, introduction of a transition layer modifies scattering and absorption of light. In radially inhomogeneous media, Eqs.~(\ref{WEd_H}) and (\ref{WEd_E}) for radial distributions change to 
\begin{eqnarray}
	&\displaystyle 
	\frac{\d^2 (rH_{lm})}{\d r^2}-\left[\frac{l(l+1)}{r^2}-k_0^2\varepsilon(r)\right](rH_{lm})=\nonumber \\
	&\displaystyle 
	\hspace{4cm}\frac{1}{\varepsilon(r)}\frac{\d \varepsilon(r)}{\d r}\frac{\d(r  H_{lm})}{\d r},\label{WEd_H_inh}\\
	&\displaystyle
		\frac{\d^2 (rE_{lm})}{\d r^2}-\left[\frac{l(l+1)}{r^2}-k_0^2\varepsilon(r)\right](rE_{lm})=0. \label{WEd_E_inh}
\end{eqnarray}
By introducing the relative TM wave impedance $\tilde Z_l$ and TE wave admittance $\tilde Y_l$ normalized with the free-space impedance $Z_0$:  
\begin{eqnarray}
	&\displaystyle 
	\tilde Z_l=\frac{Z_l}{Z_0}=- \frac{\i}{k_0\varepsilon(r)}\frac{1}{r H_{lm}}\frac{\d (r H_{lm})}{\d r},\\
	&\displaystyle
	\tilde Y_l=Z_0Y_l=\frac{\i}{k_0}\frac{1}{r E_{lm}}\frac{\d (r E_{lm})}{\d r},
\end{eqnarray}
Eqs.~(\ref{WEd_H_inh}) and (\ref{WEd_E_inh}) can be rewritten for $\tilde Z_l$ and $\tilde Y_l$:
\begin{eqnarray}
	&\displaystyle 
	\frac{\d \tilde Z_l}{\d r}-\i k_0\left[1-\varepsilon(r)\tilde Z_l^2-\frac{l(l+1)}{k_0^2r^2\varepsilon(r)}\right]=0,
\label{Z_H}\\
	&\displaystyle 
	\frac{\d \tilde Y_l}{\d r}+\i k_0\left[\varepsilon(r)-\tilde Y_l^2-\frac{l(l+1)}{k_0^2r^2}\right]=0.
\label{Y_E}
\end{eqnarray}
Solving them with $\tilde Z_l$ and $\tilde Y_l$ given by internal fields (\ref{H_int}) and (\ref{E_int}) at the inner boundary $r=R_i$ of the transition layer, we can get the distributions of $\tilde Z_l$ and $\tilde Y_l$ across the entire layer. Then, continuity of the wave impedance and admittance at the outer boundary $r=R_e$ gives us the modified Mie scattering coefficients: 
\begin{eqnarray}
	\displaystyle 
	a_{l}=\frac{q_i\tilde R_e\psi_l(q_i)\psi_l'(q_e)-q_e\tilde R_i\psi_l(q_e)\psi_l'(q_i)+F_l\psi_l(q_e)}
	{q_i\tilde R_e\psi_l(q_i)\xi_l'(q_e)-q_e \tilde R_i\xi_l(q_e)\psi_l'(q_i)+F_l\xi_l(q_e)},~~~
\label{a_l_tra}\\
	\displaystyle 
	b_{l}=\frac{q_e\tilde R_i\psi_l(q_i)\psi_l'(q_e)-q_i\tilde R_e\psi_l(q_e)\psi_l'(q_i)+G_l\psi_l(q_e)}
	{q_e\tilde R_i\psi_l(q_i)\xi_l'(q_e)-q_i\tilde R_e\xi_l(q_e)\psi_l'(q_i)+G_l\xi_l(q_e)},~~~
\label{b_l_tra}
\end{eqnarray}
with $\tilde R_i=2R_i/(R_i+R_e)$, $\tilde R_e=2R_e/(R_i+R_e)$, $q_i=k_iR_i$, $q_e=k_eR_e$ and the factors 
\begin{eqnarray}
	&\displaystyle 
	F_l=-2\i\Delta \tilde Z_l\frac{q_iq_e}{k_0(R_i+R_e)}\psi_l(q_i),\\
	&\displaystyle 
	G_l=2\i \Delta \tilde Y_l  \frac{k_0 R_e R_i}{R_i+R_e}\psi_l(q_i)
\end{eqnarray} 
accounting for the the transition layer impedance $\Delta \tilde Z_l=\tilde Z_l(R_e)-\tilde Z_l(R_i)$ and admittance $\Delta \tilde Y_l=\tilde Y_l(R_e)-\tilde Y_l(R_i)$.  

For thin layers, $\Delta \tilde Z_l$ and $\Delta \tilde Y_l$ are proportional to thickness of the transition, $\Delta R=R_e-R_i$. Therefore, effects of $\Delta \tilde Z_l$ and $\Delta \tilde Y_l$ on scattering coefficients are often weak, except the case when $\Delta \tilde Z_l$ become resonant. The latter happens when the transition layer has a point $r=r_0$ where ${\rm Re}~\varepsilon(r_0)=0$. In this case, the last term in Eqs.~(\ref{Z_H}) exhibits a complex pole that contributes
\begin{equation}
\Delta \tilde Z_l^{\rm res}=-\i \int^{r_0+0}_{r_0-0}\frac{l(l+1)}{k_0r^2\varepsilon(r)}\d r
\end{equation} 
to $\Delta \tilde Z_l$ with a non-zero real part
\begin{equation}
{\rm Re}~\Delta \tilde Z_l^{\rm res}\approx\pi\frac{l(l+1)}{k_0r_0^2}\left(\frac{\d{\rm Re}~\varepsilon}{\d r}\right)^{-1}_{r=r_0}.
\end{equation} 
Thus, presence of the resonant point $r_0$ inside the transition enhances its intrinsic absorption for the TM spherical harmonics, especially with high $l$. This, however, requires a dielectric permittivity $\varepsilon(r)$ changing sign of its real part inside the transition layer. The above condition is fulfilled for ${\rm Re}~\varepsilon_i/\varepsilon_e<0$. In such structures, the induced electric fields experience the localized resonance on longitudinal polarization eigenmodes, such as phonons, plasmons or excitons, for every orbital index $l$. The enhanced fields increase superficial absorption in the sphere and change the volume contributions to $\sigma_{\rm sca}$ and $\sigma_{\rm abs}$ \cite{Akimov:2012}. 

\begin{figure}[t]
	\includegraphics[width=0.48\textwidth,clip,trim={0.2cm 17cm 0.2cm 0.1cm}]{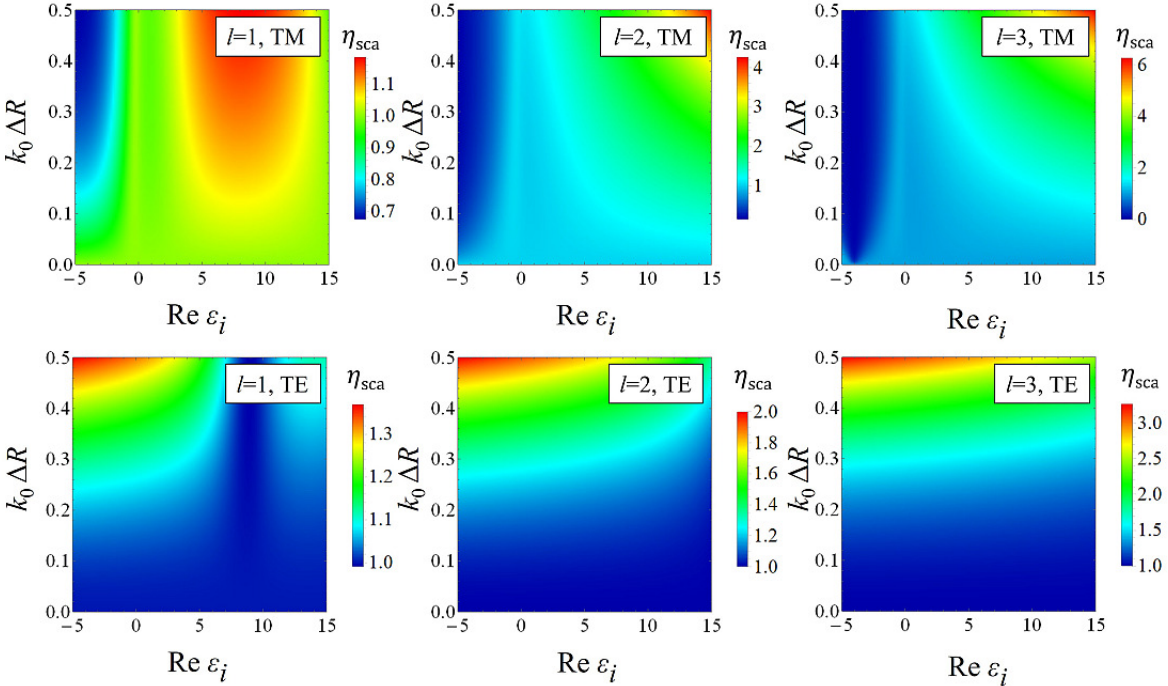}
	\caption{Ratios $\eta_{\rm sca}$ of the modified scattering cross-section contributions given by scattering coefficients (\ref{a_l_tra}) and (\ref{b_l_tra}) to the classical ones for the TM and TE scattered fields with $l=1,2,3$ under $k_0(R_i+R_e)/2=1$, $\varepsilon_e=2.25$, ${\rm Im}~\varepsilon_i=0.1$ for linear profiles of $\varepsilon(r)$.
	\label{Fig11}}
\end{figure}

\begin{figure}[t]
	\includegraphics[width=0.48\textwidth,clip,trim={0.2cm 17cm 0.2cm 0.1cm}]{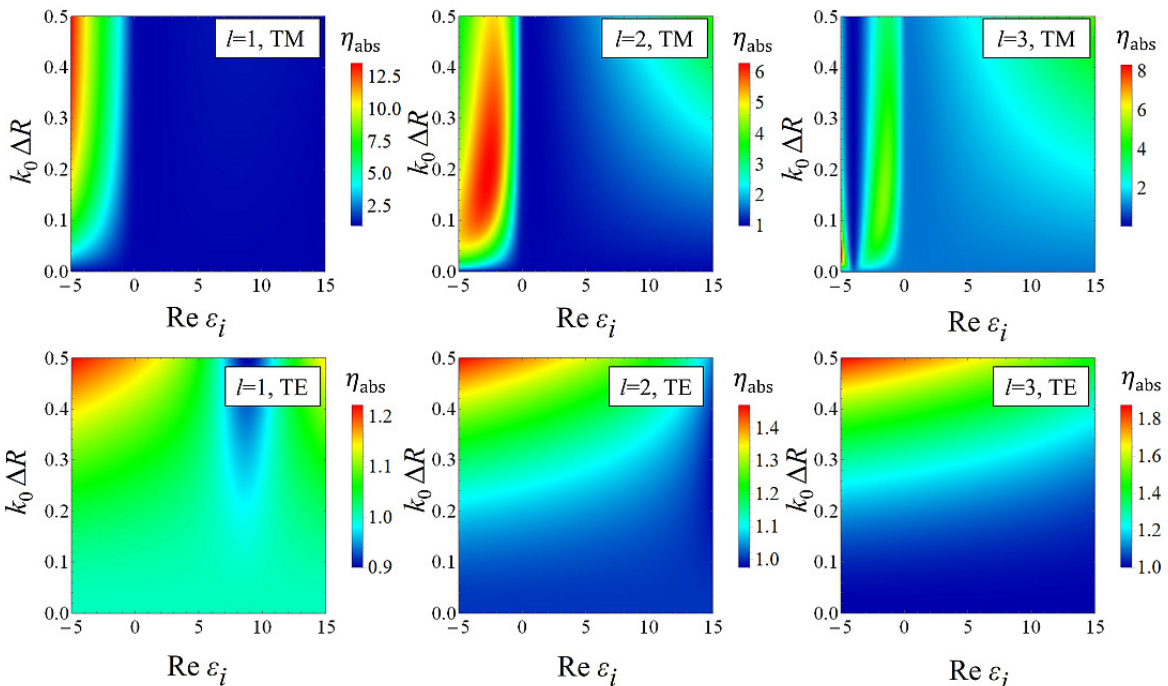}
	\caption{Ratios $\eta_{\rm abs}$ of the modified absorption cross-section contributions given by scattering coefficients Eqs.~(\ref{a_l_tra}) and (\ref{b_l_tra}) to the classical ones for the TM and TE scattered fields with $l=1,2,3$ under $k_0(R_i+R_e)/2=1$, $\varepsilon_e=2.25$, ${\rm Im}~\varepsilon_i=0.1$ for linear profiles of $\varepsilon(r)$.
	\label{Fig12}}
\end{figure}

The typical effects of transitions on the modified contributions of the TM and TE spherical harmonics to $\sigma_{\rm sca}$ and $\sigma_{\rm abs}$ are shown in Figs.~\ref{Fig11} and \ref{Fig12} for linear profiles of $\varepsilon(r)$. Governed by the surface-to-volume ratio, continuous transitions have stronger effect at larger $\Delta R$ and smaller $R_i$. 

\subsection{Scatterer localization}

Noteworthy that neither the account for a current source nor the inclusion of a transition layer considered above solves the issue of source-free scattered fields. In both the cases, the spherically outgoing part of the incident fields can still annihilate with the induced scattered fields following the trivial solution of Maxwell's equations. This annihilation unphysical from the electrodynamics point of view comes mathematically possible owing to differential equations (\ref{WEd_H}) and (\ref{WEd_E}) that govern distributions of {\it both} the incident and scattered fields in the external domain. To understand how this annihilation can be avoided, we recall the fundamental difference in definition of the incident and scattered fields. 

Conventional solution to any scattering problem consists of two steps. 
The first step is consideration of the problem {\it without} the scatterer to get the field distribution for incident radiation. 
In Mie theory, it corresponds to finding the source-free incident fields supported by a uniform external medium. 
The second step is solution of the problem {\it with} the scatterer included for the given incident fields found at the first step. 
In other words, incident and scattered fields are defined in two structurally different systems of (i) the boundless external medium and (ii) the boundless external medium with the scatterer. 
In Mie theory, these two systems have the common domain at $r>R$, where both the incident and scattered fields fulfill the same equations and, therefore, can completely compensate each other. 
To avoid this, the systems (i) and (ii) should not have any common exterior domain. 
This requires the scatterer to be extended over the entire space and vanish only at $r\rightarrow \infty$. 
This means, the scatterer should be boundless. It still can have a major localization in some part of space, but never be fully confined in any finite domain.  

In other words, the origin of source-free scattered fields observed in Mie theory is the finite localization of the scatterer at $r<R$. 
Continuous extension of the scatterer beyond $r=R$ with weak contributions at $r>R$ solves the issue of electrodynamically unphysical scattered fields. 
However, solution of the scattering problem for a radially inhomogeneous sphere significantly differs from the classical Mie theory for a piecewise homogeneous sphere, so the scattered fields can no longer be described with Eqs.~(\ref{H_sca}) and (\ref{E_sca}). 
Therefore, the classical scattering coefficients cannot be modified to account for an infinitely extended scatterer.   

\section{Conclusion}

In conclusion, we have performed the spherical harmonic analysis of the scattered fields given by Mie theory and revealed the two groups of source-free and current-sourced fields it deals with.
These fields have been shown to experience distinct dynamics and influence by the sphere.
We have demonstrated how by interfering with each other, these fields make scattering and absorption resonant and, thus, lead to all the spectral features observed in Mie theory.  
We have discussed the limitations caused by the use of source-free fields, as well as considered the ways for further refinement of the classical theory for description of the excitation source, sphere interface and scatterer localization.
The limitations and solutions discussed have been shown to be common to most of scattering problems and equally applicable to non-spherical scatterers.

\end{document}